\documentclass[showpacs,preprintnumbers,preprint]{revtex4}
\usepackage{amsmath}
\usepackage{graphicx}
\usepackage{dcolumn}
\usepackage{bm}

\usepackage{amssymb}
\usepackage{makeidx}
\usepackage{amsmath}
\usepackage{lscape}
\usepackage{epsfig}
 \newcommand{\newc}{\newcommand}
\newc{\beq}{\begin{equation}} \newc{\eeq}{\end{equation}}
\newc{\bea}{\begin{array}} \newc{\eea}{\end{array}}
\newc{\ri}{{\mathrm i}}
\newcommand{\p}{\partial}

\begin{document}

\allowdisplaybreaks
 \title { Symmetries of field equations
of axion electrodynamics}
\author{A.G. Nikitin}
\email{nikitin@imath.kiev.ua}
\affiliation{ Institute of
Mathematics, National Academy of Sciences of Ukraine,\\ 3
Tereshchenkivs'ka Street, Kyiv-4, Ukraine, 01601}
\author{ Oksana Kuriksha}
\email{kuriksha@imath.kiev.ua}
 \affiliation{Petro Mohyla Black Sea State University,\\
 10, 68 Desantnukiv Street,
 54003 Mukolaiv,
 Ukraine}
 \date{\today}
\pacs{03.65.Pm,03.65.Fd, 03.65.Ge, 03.50.Kk} \keywords{Axion
electrodynamics, group classification, conservation laws, exact
solutions}
\begin{abstract} The group classification of models of axion
electrodynamics with arbitrary self interaction of axionic field is
carried out. It is shown that  extensions of the basic Poincar\'e
invariance of these models appear only for constant and exponential
interactions. The related conservation laws are discussed. The
maximal continuous symmetries of  the 3d Chern-Simons
electrodynamics and Carroll-Field-Jackiw electrodynamics are
presented. Using the In\"on\"u-Wigner contraction the
nonrelativistic limit of equations of axion electrodynamics is
found. Exact solutions for the electromagnetic and axion fields are
discussed including those
 which describe propagation with group
velocities faster than the speed of light. However these
solutions are causal since the corresponding energy velocities are
subluminal.
\end{abstract}

\maketitle


\section{Introduction}
   To explain the absence of the CP symmetry violation in
interquark interactions Peccei and Quinn \cite{pec} suggested that a
new symmetry must be
 present. The breakdown of this gives rise to the
axion field proposed  later by Weinberg \cite{weinberg} and Wilczek
\cite{wilczek1}. And it was Wilczek who presented the first analysis
of possible effects caused by axions in electrodynamics
\cite{wilczek}. Notice that the idea to include an extra
pseudoscalar field  into electrodynamics was proposed by Ni
\cite{Ni} as early as 1974, and so this date can be treated as the
birth year of the axion prototype.

Axions belong to the main candidates to form the dark matter, see,
e.g. \cite{raflet} and references cited therein. New arguments for
the materiality of axion theories were created in solid states
physics. Namely, it was found recently \cite{Qi} that the
axionic-type interaction terms appear in the theoretical description
of a class of crystalline solids called topological insulators.
Axion electrodynamics gains plausibility by results of Heht et al
\cite{heht} who extract the existence of a pseudoscalar field from
the experimental data concerning electric field-induced
magnetization on $\text{Cr}_2\text{O}_3$ crystals or the magnetic
field-induced polarization. In other words, although their existence
is still not confirmed experimentally  axions are stipulated at
least in the three fundamental fields: QCD, cosmology and condensed
matter physics.

There are many other interesting aspects of axion electrodynamics.
In particular, its reduced version (corresponding to the external
axion field linear in independent variables) was used by by Carroll,
Field and Jackiw (CFJ) \cite{jackiw} to examine the possibility of
Lorentz and CPT violations in Maxwell's electrodynamics. In
addition, just the interaction Lagrangian of axion electrodynamics
generalizes the Chern-Simons form $\varepsilon_{abc}A^a\nabla^bA^c$
\cite{chern} to the case of (1+3)-dimensional Minkowski space.

Let us present  more arguments for materiality of axion
electrodynamics which are very inspiring for us. Recently  new
exactly solvable models for neutral Dirac fermions had been
discovered \cite{ninni}, \cite{NN4}. These models involve the
external electromagnetic fields which do not solve Maxwell equations
with physically reasonable currents. However, these fields solve
equations of axion electrodynamics. We had classified exactly
solvable quantum mechanical models with matrix potentials
\cite{Nik2}, \cite{NKa2}, and superintegrable models of cold
neutrons \cite{N3}. Some of these systems also include external
fields which solve equations of axion electrodynamics. In addition,
these field equations appear to be a relativistic counterpart of
Galilei invariant systems classified in \cite{NN3}. Thus we have a
particular interest to study equations of axion electrodynamics, and
we will do it using the tools of group theory.

Group theory, and especially the theory of Lie groups is one of the
corner stones of  modern theoretical physics. Symmetries of
Lagrangians and of the corresponding motion equations form a very
essential constituent part of any physical theory. However, except
the analysis of symmetries of the CFJ model presented in paper
\cite{har}, we do not know any systematical investigation of
symmetries of axion theories. Notice that such an investigation
would generate group-theoretical backgrounds for axion models and
enable to construct their exact solutions.

In the present paper we make the group classification of the field
equations of axion electrodynamics with arbitrary self interaction
of axion field. The considered model includes the standard axion
electrodynamics as a particular case.  We prove that an extension of
the basic Poincar\'e invariance appears only for the exponential,
constant and trivial interaction terms. These and other results of
group classification are presented in Section 3 and Appendix A.

In addition, we carry out the group analysis of two other theories
which are close to axion electrodynamics. Namely, we describe Lie
symmetries  of the field equations of classical electrodynamics
modified by adding the Chern-Simons term, and symmetries of the CFJ
model. As it is shown in Appendix B, the maximal continuous group of
Chern-Simon electrodynamics is the 17-parametrical extended
conformal group.

A special subject of our analysis are conservation laws which
correspond to found symmetries. They are discussed in Section 4,
 where we present a simple proof that the interaction
 between the electromagnetic and axion fields
  does not affect the energy-momentum tensor.

  In Section 5 we present selected invariant solutions of field equations
   of axion electrodynamics. Some of these solutions play the key
role in formulation of exactly solvable problems of quantum
mechanics in both relativistic \cite{ninni}, \cite{N4} and
nonrelativistic \cite{N3}, \cite{Pron} approaches.

In Section 6 we analyze  plane wave solutions which are smooth and
bounded functions which generate positive definite and bounded
energy density. We show that these solutions describe waves whose
group velocity can be superluminal.  Nevertheless, they are causal
since the corresponding energy velocities  are smaller than the
velocity of light.

An important constituent of any relativistic model is its
nonrelativistic limit. This is true also for models including
massless fields.  As it was shown long time ago \cite{lebellac},
there exist a reasonable (and very important) nonrelativistic
approximation for the Maxwell equations, which makes them invariant
w.r.t. the Galilei group. Namely, in this approximation we obtain
equations of Faraday electrodynamics. This result justifies Galilei
invariance
 of  quantum mechanical systems including  particles interacting with an external electromagnetic field.

A natural question arises whether it is possible to extend this
result to the case of field equations of axion electrodynamics.
 Notice that the correct definition of the
nonrelativistic limit of a physical model is by no means a simple
problem in general and in the case of theories of massless fields in
particular, see, for example, \cite{Hol}.   Such limit is not
necessary unique, and simple passing the speed of light to infinity
we can obtain a physically meaningless theory.

By definition, any relativistic system is invariant w.r.t. the
Poincar\'e group P(1,3), and a correct nonrelativistic approximation
of this system should be invariant w.r.t. the Galilei group G(1,3).
Thus to obtain a well defined nonrelativistic limit it is necessary
to take a care on the attending transformation P(1,3)$\to$G(1,3).
This idea had been proposed long time ago by In\"on\"u and Wigner
\cite{contraction} who presented definitions and justifications for
such transformation. It is a special limiting procedure called
contraction, which is an important subject of modern group theory.

 In Section 7 we  find a
nonrelativistic limit of equations of  axion electrodynamics with
using a generalized In\"on\"u-Wigner (IW) contraction. As a result
we prove that the Galilei-invariant wave equations for an abstract
ten-component vector field, deduced in \cite{NN3}, are nothing but a
contracted version of the field equations of axion electrodynamics.

Appendix A includes a rather detailed proof of the results
formulated in Section 3. Finally, in Appendix B we present the
results of group analysis of the field equations of classical
electrodynamics modified by adding the Chern-Simons terms, and of
the CFJ model.

\section{Field equations of axion electrodynamics}
Let us start with the following model Lagrangian:
\begin{eqnarray}\label{1}L=\frac12 p_\mu p^\mu-\frac14 F_{\mu\nu}F^{\mu\nu}+
\frac{\kappa}{4}\theta F_{\mu\nu}\widetilde
F^{\mu\nu}-V(\theta).\end{eqnarray} Here $F_{\mu\nu}$ is the
strength tensor of electromagnetic field, $\widetilde
F_{\mu\nu}=\frac12 \varepsilon_{\mu\nu\rho\sigma}F^{\rho\sigma}$,
$p_\mu=\partial_\mu\theta$, $\theta$ is the pseudoscalar axion
field, $V(\theta) $ is a function of $\theta $, $\kappa$ is a
dimensionless constant, and the summation is imposed over the
repeating indices over the values 0, 1, 2, 3. Moreover, the strength
tensor can be expressed via four-potential $A=(A^0, A^1, A^2, A^3)$
as:
\begin{eqnarray}F^{\mu\nu}=\p^\mu A^\nu-\p^\nu
A^\mu.\label{pot}\end{eqnarray}

Setting in (\ref{1}) $\theta=0$ we obtain the Lagrangian for Maxwell
field. Moreover, if $\theta$ is a constant then (\ref{1}) coincides
with the Maxwell Lagrangian up to constant and four-divergence
terms. Finally, the choice $V(\theta)=\frac12m^2\theta^2$ reduces
$L$ to the standard Lagrangian of axion electrodynamics.

We will investigate symmetries of the generalized Lagrangian
(\ref{1}) with  arbitrary $V(\theta)$. More exactly, we will make
the group classification of the corresponding Euler-Lagrange
equations:
\begin{eqnarray}&&\partial_\nu F^{\mu\nu}=\kappa
p_\nu{\widetilde F}^{\mu\nu},\label{short1}\\&&
\label{short2}\partial_\nu \partial^\nu\theta=-\frac{\kappa}{2}
F_{\mu\nu}\widetilde F^{\mu\nu}+F
\end{eqnarray}
where $F=-\frac{\p V}{\p\theta}$. In addition, in accordance with
its definition, $\widetilde F^{\mu\nu}$ satisfies the Bianchi
identity
\begin{eqnarray}\p_\nu\widetilde F^{\mu\nu}=0\label{short3}.\end{eqnarray}

Substituting (\ref{pot}) into (\ref{short1}) one obtains the second
order equation for potential $A_\mu$:
\begin{eqnarray}\partial_\nu \partial^\nu A^\mu=-\kappa
p_\nu{\widetilde F}^{\mu\nu}\label{amu}\end{eqnarray} provided
$A_\mu$ satisfies the Lorentz gauge condition:
\begin{eqnarray}\label{lg}\partial_\mu A^\mu=0.\end{eqnarray}

 Just the system of equations
(\ref{short1})--(\ref{short3}) will be the main subject of group
classification. In addition, we shall discuss symmetries of field
equations of the Chern-Simons electrodynamics, i.e., of the system
including equations (\ref{short1}) and (\ref{short3}). In this
theory $p_\mu$ is treated as an external field whose motion equation
is not specified, i.e., equation (\ref{short2}) is ignored.

\section{Group classification of systems
(\ref{short1})--(\ref{short3})}

Equations (\ref{short1})--(\ref{short3}) include an arbitrary
function $F(\theta)$ so we can expect that  the variety of
symmetries of this system depends on the explicit form of $F$. The
group classification of these equations presupposes finding their
symmetry groups for arbitrary $F$.

In this section we present the results of group classification while
the related calculations details are given in Appendix A.

The maximal continuous symmetry of system
(\ref{short1})--(\ref{short3}) with arbitrary function $F(\theta)$
is given by Poincar\'e group $P(1,3)$. On the set of solutions of
equations (\ref{short1})--(\ref{short3}) written as a seven
component vector
\begin{eqnarray}\label{cal}{\cal F}=\text{column}(F^{01}, \ F^{02}, \
F^{03},\ F^{23}, \ F^{31},\ F^{12},\ \theta)\end{eqnarray}infinitesimal
generators of this group take the following form:
\begin{eqnarray}
P_\mu=\p_\mu,   \quad J_{\mu\nu}=x_\mu
\p_{\nu}-x_\nu\p_{\mu}+S_{\mu\nu}
\label{kuriksha:yadro_operators}\end{eqnarray} where indices $\mu$
and $\nu$ independently take values 0, 1, 2, 3,
\[S_{ab}=\varepsilon_{abc}\begin{pmatrix}S_c&\cdot&\cdot\\
\cdot&S_c&\cdot\\\cdot&\cdot&0\end{pmatrix},\quad S_{0c}=
\begin{pmatrix}\cdot&-S_c&\cdot\\
S_c&\cdot&\cdot\\\cdot&\cdot&0\end{pmatrix},\quad a, b, c\neq0,\]
$S_c$ are $3\times3$ matrices of spin 1, whose entries are
$(S_c)_{ab}=\text{i}\varepsilon_{cab}$,  $\varepsilon_{cab}$ is the
Levi-Chivita symbol, and the dots denote the zero matrices of an
appropriate dimension.

Alternatively, spin matrices $S_{\mu\nu}$ can be represented as the
first order differential operators:
\begin{eqnarray}\label{smunu}S_{\mu\nu}= F_{\mu\lambda}
\p_{F^{\lambda\nu}}-F_{\nu\lambda}
\p_{F^{\lambda\mu}}.\end{eqnarray}
Such notation is both convenient and usual for group analysis of differential equations.

Operators (\ref{kuriksha:yadro_operators}) form a basis of the Lie
algebra p(1,3) of the Poincar\'e group.

For some special functions $F(\theta)$ symmetry of system
(\ref{short1})--(\ref{short3}) appears to be more extended. Namely,
if $F=0$, $F=c$ or $F=b\exp(a\theta)$ then the basis
(\ref{kuriksha:yadro_operators}) of symmetry algebra of this system
is extended by the following additional operators $P_4,\ D$ and $X$:
\begin{eqnarray}\begin{matrix}P_4=\p_{\theta}, \quad D=x_0 \p_0+x_i\p_{i}-
\frac12F^{\mu\nu}\p_{F^{\mu\nu}}
& \text{ if }\  F(\theta)=0,\\
P_4=\p_{\theta}&  \text{ if } \ F(\theta)=c, \\
X=aD-2P_4& \text{ if }\ F(\theta)=b \texttt{e}^{a
\theta}.\end{matrix}\label{kuriksha:rozsh_operators3}
\end{eqnarray}
Operator $P_4$ generates shifts of dependent variable $\theta$, $D$
is the dilatation operator generating a consistent scaling of
dependent and independent variables, and $X$ generates the
simultaneous shift and scaling. Note that arbitrary parameters $a,
b$ and $c$ can be reduced to the fixed values $a=\pm1$, $b=\pm1$ and
$c=\pm1$ by scaling dependent and independent variables.

 Thus the continues symmetries of
system (\ref{short1})--(\ref{short3}) where $F(\theta)$ is an
arbitrary function of $\theta$ are exhausted by the Poincar\'e
group. The same symmetry is accepted by the standard equations of
axion electrodynamics which correspond to $F(\theta)=-m^2\theta$. In
the cases indicated in (\ref{kuriksha:rozsh_operators3}) we have the
extended Poincar\'e groups.

Notice that symmetries of the equations (\ref{amu}), (\ref{lg}),
(\ref{short3}) for potentials also are generated by infinitesimal
operators of the form (\ref{kuriksha:yadro_operators}) where
\begin{eqnarray}\label{sa}S_{\mu\nu}=A_\mu\p_{A_\nu}-A_\nu\p_{A_\mu}.\end{eqnarray}
Additional symmetries again are given by equations
(\ref{kuriksha:rozsh_operators3}) where, however,  $D\to x^\mu \p_\mu-
A^{\mu}\p_{A^{\mu}}.$
\section{Conservation laws}

An immediate consequence of  symmetries presented above is the
existence of conservation laws. Indeed, the system
(\ref{short1})--(\ref{short3}) admits a Lagrangian formulation.
Thus, in accordance with the Noether theorem, symmetries of
equations (\ref{short1})--(\ref{short3}) which keep the shape of
Lagrangian (\ref{1}) up to four divergence terms should generate
conservation laws. Let us present them
 explicitly.

First we represent  generators (\ref{kuriksha:yadro_operators}),
(\ref{sa}) and (\ref{kuriksha:rozsh_operators3}) written in terms of
the variational variables $A^\mu$ and $A^4=\theta$ in the following
unified form:
\begin{eqnarray}
Q= \xi^\mu\p_{\mu}+\varphi^{\tau} \p_{A^\tau} \label{pota}
\end{eqnarray}
where the summation is imposed over the values $\tau=0,1,2,3,4$ and
$\mu=0,1,2,3$.

Conserved current corresponding to symmetry (\ref{pota}) can be
represented as \cite{olver}:
\begin{eqnarray}J_\sigma=\varphi_\tau\frac{\p L}{\p (\p_\sigma
A_\tau)}+\xi^\sigma L-\xi^\nu \p_\nu A^\tau\frac{\p L}{\p (\p_\sigma
A^\tau)}\label{cv}.\end{eqnarray}

  The basic conserved quantity is
the energy-momentum tensor which corresponds to symmetries $P_\mu$
presented in (\ref{kuriksha:yadro_operators}). In this case
\begin{eqnarray}\label{Pmu}\varphi^\tau\equiv0\quad  \texttt{and} \quad \xi_\mu=1\end{eqnarray} where $\mu$   successively takes the values  0, 1, 2, 3 .  Substituting  (\ref{1}), (\ref{Pmu}) into (\ref{cv})  and
using three dimensional notations
\begin{eqnarray}F_{0a}=E_a,\quad F_{ab}=\varepsilon_{abc} B_c.
\label{s2}\end{eqnarray}
 we find the components of the conserved energy-momentum tensor  in the following form:
\begin{eqnarray}\begin{aligned}\label{cl1}&T^{00}=\frac12(\mathbf{E}^2+\mathbf{B}^2+
p_0^2+\mathbf{p}^2)+V(\theta),\ \
T^{0a}=T^{a0}=\varepsilon_{abc}E_bB_c+p^0p^a,\\&
T^{ab}=-E^aE^b-B^aB^b+p^ap^b+\frac12\delta^{ab}(\mathbf{E}^2+
\mathbf{B}^2+p_0^2-\mathbf{p}^2-2V(\theta)).
\end{aligned}\end{eqnarray}

  The tensor $T^{\mu\nu}$ is symmetric and satisfies the continuity equation
 $\partial_\nu T^{\mu\nu}=0$. Its components $T^{00}$ and
 $T^{0a}$ are associated with the energy and momentum densities.

 It is important to note that the
 energy-momentum  tensor does not depend on
parameter $\kappa$ and so is not affected by the term
$\frac{\kappa}{4}\theta F_{\mu\nu}\widetilde F^{\mu\nu}$ present in
Lagrangian (\ref{1}). In fact this tensor is nothing but a sum of
energy momenta  tensors for the free electromagnetic field and
scalar field. Moreover, the interaction of these fields between
themselves is not represented in (\ref{cl1}).

The conservation of tensor (\ref{cl1}) is caused by the symmetry of
Lagrangian (\ref{1}) w.r.t. shifts of independent variables $x_\mu$.
The symmetries w.r.t. rotations and Lorentz transformations give
rise to conserving of the following tensor:
\begin{eqnarray}\label{oks}G^{\alpha\nu\mu}=x^\alpha T^{\mu\nu}-x^\nu
T^{\mu\alpha}\end{eqnarray} which satisfies the continuity equation
w.r.t. the index $\mu$. In particular, for $\alpha, \nu=1,2,3$
equation (\ref{oks}) represents the angular momentum tensor.

The tensors (\ref{cl1})--(\ref{oks}) exhaust the conserved
quantities whose existence is caused by the Lie symmetries of
equations (\ref{short1})--(\ref{short3}) with arbitrary function
$F(\theta)$.

The additional symmetries presented in
(\ref{kuriksha:rozsh_operators3}) are neither variational nor
divergent symmetries of Lagrangian (\ref{1}), and so they do not
generate conservation laws. However, we can indicate another
conservation laws which have nothing to do with Lie symmetries.

First let us  note that equation (\ref{short1})  in itself can be written in
the divergence form $\p_\nu {\cal F}^{\mu\nu}=0$, where
\beq\nonumber{\cal F}^{\mu\nu}=F^{\mu\nu}-\kappa\theta\tilde
F^{\mu\nu}\eeq is the antisymmetric conserved tensor. In addition,
this equation
 can be
rewritten as $j_\mu=0$ where \beq\label{cc}j^\mu=\partial_\nu
F^{\mu\nu}-\kappa p_\nu{\widetilde F}^{\mu\nu}\eeq is a conserved
current. Changing equation (\ref{short1}) by  (\ref{cc}) with
$j^\mu\neq0$ we obtain the system which represents the field
equations of
 axion electrodynamics with  nontrivial
currents.

Equation (\ref{short2}) in its turn  can be represented as
\begin{eqnarray}\p_\mu J^\mu=F(\theta)\nonumber\end{eqnarray}
 where
\begin{eqnarray}\label{ce}J^\mu=p^\mu+
\kappa\widetilde F^{\mu\nu}A_\nu.\end{eqnarray} If $F=0$ then
current (\ref{ce}) satisfies the continuity equation.

In addition to (\ref{cl1})--(\ref{ce}) there exist the infinite
number of (trivial) conserved currents corresponding to the gauge
symmetries of Lagrangian (\ref{1}).  An example of  such conserved
current is:
\begin{eqnarray}\label{ok}N^\mu=F^{\mu\nu} p_\nu \varphi(\theta)
\end{eqnarray}
where $\varphi(\theta)$ is an arbitrary differentiable function of
$\theta$. Vector $N^\mu$ satisfies the continuity equation $\p_\mu
N^\mu=0$ provided equations (\ref{short1}) are satisfied (remember
that $p_\nu=\partial_\nu\theta$).

\section{Selected exact solutions}

 The field equations
of axion electrodynamics  form a rather complicated system of
nonlinear partial differential equations. However, this system
admits an extended symmetry algebra which makes it possible to find
a number of exact solutions. Here we present some of these solutions
while the completed list of them can be found in
 \cite{kura}.

The algorithm for construction of group solutions of partial
differential equations goes back to Sophus Lie and is expounded in
various monographs, see, e.g., \cite{olver}. Roughly speaking, to
find such solutions we have to change the dependent and independent
variables by invariants of the subgroups of our equations symmetry
group. Solving equations (\ref{short1})--(\ref{short3}) it is
reasonable to restrict ourself  to three-parametrical subgroups of
P(1,3) which enables  to reduce
(\ref{short1})--(\ref{short3}) to systems of {\it ordinary}
differential equations. The complete list of these  subgroups can be
found in \cite{patera}.

 To make  solutions of equations (\ref{short1})--(\ref{short3})
 more physically transparent, we write them in terms
 of electric field ${\bf E}$ and magnetic field $\bf B$ whose components
  are expressed via the strengths tensor $F_{\mu\nu}$ as shown in (\ref{s2}).
In addition, we rescale the dependent  variables such  that
$\kappa\to1$.
 \subsection{Plane wave solutions}
 Let us present solutions of system (\ref{short1})--(\ref{short3})  which are invariant
 w.r.t. subalgebras of p(1,3) whose basis elements have  the
 following unified form: $\langle P_1,P_2,k P_0+\varepsilon P_3\rangle $ where
 $\varepsilon$ and $k$ are
 parameters satisfying
  $\varepsilon^2\neq k^2$, while $P_1, \ P_2,\ P_3$ and $P_0$ are generators given
  in
  (\ref{1}).

  The invariants $\omega$ of the corresponding three-parametrical group
  should solve the equations
\begin{eqnarray}\label{inv}P_1\omega=0,\quad P_2\omega=0,\quad (k P_0+\varepsilon
P_3)\omega=0.\end{eqnarray}
 Solutions of (\ref{inv}) include all dependent variables
 $E_a, B_a, \theta$ ($a=1,2,3$) and the only independent variable
 $\omega=\varepsilon x_0
-kx_3$. Thus we can search for solutions which are functions of
$\omega$ only. As a result we reduce equations
(\ref{short1})--(\ref{short3}), (\ref{s2}) to the  system of
ordinary differential equations whose solutions are
\begin{eqnarray}\begin{aligned}\label{bed}&B_1=-kc_1\theta,\
 \  B_2=\varepsilon c_1+kc_2,\ \
B_3=c_3,\\& E_1=\varepsilon c_2 +kc_1, \ \ E_2=\varepsilon c_1\theta
,\ \ E_3=c_3\theta-c_4(\varepsilon^2-k^2)\end{aligned}\end{eqnarray}
where $c_1,...,c_4$ are arbitrary real numbers. The corresponding
bounded  solutions of equation (\ref{short2}) with $F=-m^2\theta$
are:
\begin{eqnarray}\label{A1a}\theta=a_\mu \cos \mu\omega+r_\mu
\sin\mu\omega +\frac{c_3c_4}{\mu^2} \end{eqnarray} where $a_\mu,
r_\mu$ and $\mu$ are arbitrary constants restricted by the following
constraint:
\begin{eqnarray}\label{mu}\mu^2=\left(c_1^2+\frac{c_3^2+m^2}{\varepsilon^2-
k^2}\right).\end{eqnarray} Notice that for the simplest nonlinear
function $F=\lambda \theta^2$  equation (\ref{short2}) is reduced to
Weierstrass one and admits a nice soliton-like solution
\beq\label{N4}
\theta={\frac{c_3c_4}{2}}\tanh^2\left(\omega+C\right)\eeq where $C$
is an integration constant. The related parameters $\varepsilon,\ k$
and $\lambda$ should satisfy the conditions
\begin{eqnarray}\label{DR2}
\varepsilon^2=k^2+\frac{ c_3^2}{8-c_1^2}, \quad \lambda
c_3c_4=12.\end{eqnarray}
 The corresponding
magnetic, electric and axion fields are localized waves moving along
the third coordinate axis.

In analogous (but as a rule much more complicated) way we can find
solutions corresponding to the other three dimensional subalgebras
of the Poincar\'e algebra. One more and rather specific solution of
equations (\ref{short1})--(\ref{short3}) with $\kappa=1$ and $F=0$
(obtained with using the subalgebra spanned on basis elements $
\langle J_{12}+k P_0+\varepsilon P_1, P_2, P_3\rangle$) can be
written as follows:
\begin{eqnarray}\label{wond1}\begin{aligned}&E_1=\varepsilon(c_k\sin(\omega)
-d_k\cos(\omega)), \quad
E_2=\varepsilon(c_k\cos(\omega)+d_k\sin(\omega)),\quad E_3=e,
\\& B_1=-\frac{k}{\varepsilon}E_2,\quad
B_2=\frac{k}{\varepsilon}E_1,\quad B_3=0,\quad \theta=\alpha x_0+\nu
x_3+\mu\end{aligned}\end{eqnarray} where $e, c_k, d_k, \varepsilon,
k, \alpha, \nu, \mu$ are constants satisfying the following
conditions:
\begin{eqnarray}\label{wond2}\varepsilon^2-k^2=\nu\varepsilon-
\alpha k,\quad \varepsilon\neq0.\end{eqnarray}

Solutions (\ref{wond1})  depend on  two different plane wave
variables, i.e., $\omega=\varepsilon x_0-kx_1$ and $\alpha x_0+\nu
x_1$. They satisfy the superposition principle since a sum of
solutions with different $\varepsilon, k, c_k$ and $d_k$ is also a
solution of equations (\ref{short1})--(\ref{short3}) with $\kappa=1$
and $F=0$. Thus it is possible to generate much more general
solutions by summing up  functions (\ref{wond1}) over $k$ and
treating $c_k$ and $d_k$ as arbitrary functions of $k$.
\subsection{Radial and planar solutions}
Consider solutions which include the Coulomb electric field. They
can be  obtained  using invariants of the subalgebra spanned on
$\langle J_{12},\ J_{23},\ J_{31}\rangle$ and have the following
form:
\begin{eqnarray}\label{solu2}
{B}_a=\frac{c_1{x}_a}{r^3}, \quad {E}_a=\frac{(c_1 \theta-c_2)
{x}_a}{r^3},\ \ \theta=\frac{\varphi}{r},
\end{eqnarray}
where $\varphi$ is a function of $x_0$ and
$r=\sqrt{x_1^2+x_2^2+x_3^2}$ satisfying the following equation:
\begin{eqnarray}\label{LL}
\frac{\partial^2\varphi}{\partial
r^2}-\frac{\partial^2\varphi}{\partial
x_0^2}=\left(\frac{c_1^2}{r^4}+m^2\right)\varphi-\frac{c_1c_2}{r^3}.
\end{eqnarray}

Setting in (\ref{solu2}) $c_1=0$ we come to the electric field of
point charge which is well defined for $r>0$. A particular solution
for (\ref{LL}) corresponding to $c_1=-q^2<0$ and $c_2=0$ is
$\varphi=c_3r\sin(mx_0)\text{e}^{-\frac{q^2}{r}}$ which gives rise
to the following field components:
\begin{eqnarray}\label{ha}{B}_a=-\frac{q^2{x}_a}{r^3}, \quad {E}_a=-\frac{q^2 \theta {x}_a}{r^3},\ \
\theta=c_3\sin(mx_0)\texttt{e}^{-\frac{q^2}{r}}.
\end{eqnarray} The components of magnetic field $B_a$ are
singular at $r=0$ while $E_a$ and $\theta$ are bounded for $0\leq
r\leq\infty$.

Separating variables it is possible to find the general solution of
equation (\ref{LL}), see \cite{kura}.

One more solution of equations (\ref{short1})-(\ref{short3}) for
$F=0$ with a radial electric field is:
\begin{eqnarray}\label{Ea}E_a=\frac{x_a}{r^2}.\end{eqnarray}
The corresponding magnetic and axion fields take the following
forms:
\[ B_1=\frac{x_1x_3}{r^2x},\
B_2=\frac{x_2x_3}{r^2x},\ B_3=-\frac{x}{r^2},\
\theta=\arctan\left(\frac{x}{x_3}\right)
\] where $x=\sqrt{x_1^2+x_2^2}.$

The electric field (\ref{Ea}) is requested in the superintegrable
model with Fock symmetry proposed in \cite{N4}.

 Let us present planar solutions which depend on spatial variables $x_1$
 and $x_2$. Namely, the functions
\begin{eqnarray} \label{N11}&&\begin{aligned}&  E_1=x_1\left({c_1}{x^{c_3-2}}
+{c_2}{x^{-2-c_3}}\right),\quad E_2=x_2\left({c_1}{x^{c_3-2}}
+{c_2}{x^{-2-c_3}}\right),\quad E_3=0,\\&
B_1=x_2\left({c_1}{x^{c_3-2}}-{c_2}{x^{-2-c_3}}\right), \quad
B_2=x_1\left({c_2}{x^{-2-c_3}}-{c_1}{x^{c_3-2}}\right),\quad
 B_3=0,\end{aligned}\\\label{N11a}
&&\ \theta=c_3\arctan\frac{x_2}{x_1}+c_4\end{eqnarray} where
$c_1,..., c_4$ are arbitrary parameters, solve equations
(\ref{short1})--(\ref{short3}) with $\kappa=1$ and $F=0$.

In particular, for $c_2=0$,  $c_3=1$ and $ c_2=0, c_3=-1$ we have:
\begin{eqnarray}\label{1/r} E_1=-B_2=\frac{c_1x_1}{x}\quad
B_1=E_2=\frac{c_1x_2}{x}, \quad
 B_3=E_3=0,\quad \theta=\arctan\frac{x_2}{x_1}\end{eqnarray}
 and
 \begin{eqnarray}\label{1/r3} E_1=-B_2=\frac{c_1x_1}{x^3}\quad
B_1=E_2=\frac{c_1x_2}{x^3}, \quad
 B_3=E_3=0,\quad \theta=\arctan\frac{x_2}{x_1}.\end{eqnarray}

Solutions (\ref{N11}), (\ref{N11a}) can be found with using
invariants of a subgroup of the {\it extended} Poincar\'e group
whose Lie algebra is spanned on the basis $\langle P_0,\ P_3,\
J_{12}+P_4\rangle $, see equations (\ref{kuriksha:yadro_operators}),
(\ref{kuriksha:rozsh_operators3}) for definitions.

Let us present an example of solutions describing fields in  a
constantly charged space. The related equation (\ref{short1}) for
$\mu=0$ should be changed to
\[\nabla\cdot{\bf  E}={\bf p}\cdot {\bf B}+j_0\]
(compare with (\ref{ss2})) where $j_0$ represents a constant charge
density.  The remaining equations (\ref{short1})--(\ref{short3}) are
kept uncharged, and the considered system is solved by the following
fields: \begin{eqnarray}\label{log}\begin{aligned}&
E_1=\frac{1}{2}j_0x_1\ln x, \quad E_2=\frac{1}{2}j_0x_2\ln x,\quad
E_3=B_3=0,\\&B_1=\frac{1}{2}j_0x_2\left(\ln x-\frac14\right),\quad
B_2=-\frac{1}{2}j_0x_1\left(\ln
x-\frac14\right),\\&\theta={2}\arctan
\frac{x_2}{x_1}.\end{aligned}\end{eqnarray}

The fields (\ref{1/r}) and (\ref{log}) appears in superintegrable
models for particles with spin 1 and $\frac32$ correspondingly
\cite{Pron} while the fields (\ref{1/r3}) are requested in the
exactly solvable system described by the Dirac equation
\cite{ninni}.

\section{Phase, group and energy velocities}

In this section we consider  some of the found  solutions in more
detail and discuss the propagation velocities of the corresponding
fields. There are various notions of field velocities, see, e.g.,
\cite{Bril} \cite{smith}, \cite{bloch}. We shall discuss the phase,
group and energy velocities.

Let us start with the plane wave solutions given by equations
(\ref{bed}) and (\ref{A1a}). They describe oscillating waves moving
along the third coordinate axis. Setting for simplicity
$c_2=c_3=c_4=r_\mu=0$ we obtain:
\begin{eqnarray}\label{ad1}\begin{aligned}&B_1=c_1 k\theta,\quad B_2=-c_1\varepsilon, \quad
B_3=0,\\&E_1=-c_1 k, \quad E_2=-c_1\varepsilon\theta,\quad
E_3=0,\quad \theta= a_\mu\cos(\mu(\varepsilon x_0-k
x_3)).\end{aligned}\end{eqnarray} Here $\varepsilon,\ k, $ and
$a_\mu$ are arbitrary parameters which, in accordance with
(\ref{mu}), should satisfy the
 the following dispersion
relations:
\begin{eqnarray}(\varepsilon^2-k^2)({\mu^2-c_1^2})={m^2}.
\label{OROR}\end{eqnarray}

If $m\neq0$ the version $\mu^2=c_1^2$ is forbidden, and we have two
qualitatively different possibilities: $\mu^2>c_1^2$ and
$\mu^2<c_1^2$.

Let $\mu^2>c_1^2$ then
$(\varepsilon^2-k^2)=\frac{m^2}{\mu^2-c_1^2}>0$. The corresponding
group velocity
  $V_g$ is
equal to the derivation of $\varepsilon$ w.r.t. $k$, i.e.,
\begin{eqnarray}\label{pease}V_g=\frac{\p\varepsilon}{\p k}=\frac{k}{\varepsilon}. \end{eqnarray}
Since $\varepsilon>k$, the group velocity appears to be less than
the velocity of light (remember that we use the Heaviside units in
which the velocity of light is equal to 1).

On the other hand the phase velocity $V_p=\frac{\varepsilon}k$ is
larger than the velocity of light, but this situation is rather
typical in relativistic field theories.

In the case $\mu^2<c_1^2$ the wave number $k$ is larger than
$\varepsilon$. As a result the group velocity (\ref{pease}) exceeds
the velocity of light, and we have a phenomenon of superluminal
motion. To understand wether the considered solutions are causal let
us calculate the energy velocity which is equal to the momentum
density divided  by the energy density:
\begin{eqnarray}V_e=\frac{T^{03}}{T^{00}}.\label{Ve}\end{eqnarray}
Substituting (\ref{ad1}) into (\ref{cl1}) we find the following
expressions for $T^{00}$ and $T^{03}$:
\[T^{00}=\frac12(\varepsilon^2+k^2)\Phi+\frac 12m^2\theta^2,\quad
T^{03}=\varepsilon k\Phi\] where
$\Phi=c_1^2(\theta^2+1)+\mu^2(a_\mu^2-\theta^2)$. Thus
\begin{gather*}V_e=\frac{2\varepsilon
k\Phi}{(\varepsilon^2+k^2)\Phi+\frac12m^2\theta^2}<\frac{2\varepsilon
k}{\varepsilon^2+k^2}<1,\end{gather*} and this relation is valid for
$\varepsilon>k$ and for $\varepsilon<k$ as well.

We see that the energy velocity is less than the velocity of light.
Thus solutions (\ref{ad1}) can be treated as causal in spite of the
fact that for $\mu^2<c_1^2$ the group velocity is superluminal.

Analogously, analyzing dispersion relations (\ref{wond2}) we
conclude, that the electromagnetic fields (\ref{wond2}) propagate
along the third coordinate axis with the group velocity
\begin{eqnarray}V_g=\left|\frac{\p\varepsilon}{\p
k}\right|=\frac1{\sqrt{1+\delta}}\label{grv}\end{eqnarray} were
\begin{eqnarray}\label{del}\delta=2\frac{\nu^2-\alpha^2}{(2k-\alpha)^2}.\end{eqnarray}

If $-1<\delta<0$ the group velocity (\ref{del}) is lager than the
velocity of light. However, the corresponding energy velocity
(\ref{Ve}) which is equal to
\begin{eqnarray}\label{dell}V_e=\frac{2\varepsilon
k+2\nu\alpha}{\varepsilon^2+k^2+\nu^2+\alpha^2}\end{eqnarray} cannot
exceed the velocity of light since $2\varepsilon
k\leq\varepsilon^2+k^2$ and $2\nu\alpha\leq\nu^2+\alpha^2$.

Thus solutions (\ref{wond1}) also can propagate with a superluminal
group velocity. However, these solutions are causal since their
energy velocities (\ref{dell}) are subluminal. Analogous results can
be proven for soliton-like solutions (\ref{N4}).

\section{nonrelativistic limit}

  To find a nonrelativistic limit of the field equations of axion electrodynamics we shall use
  the generalized IW contraction \cite{contraction} which guaranties
  Galilean symmetry of the limiting theory.

  Let us consider more
  general field
  equations including currents (see (\ref{cc})), but restrict
  ourselves  to the limiting case of the zero axionic mass. More explicitly,
  we start with the following system:
\begin{eqnarray}&&
\label{ss2}\nabla\cdot{\bf E}=\kappa{\bf p}\cdot{\bf B}+j_0,\\&&
\label{ss1}\partial_0 {\bf E}-\nabla\times{\bf B}=\kappa( p_0{\bf
B}+{\bf p}\times {\bf E})+\mathbf{j},\\&& \label{ss3}\partial_0 {\bf
B}+\nabla\times{\bf E}=0,\\&& \label{ss4}\nabla\cdot {\bf B}=0,
 \\&&
\label{ss5}\partial_0{ p_0}-\nabla\cdot{\bf p}=-\kappa{\bf
E}\cdot{\bf B}+j_4,\\&& \label{ss6}\partial_0{\bf p}-\nabla p_0=0,
\\&& \label{ss7} \nabla\times{\bf p}=0.
 \end{eqnarray}

If   $j_0=j_4=0$ and ${\bf j}=0$ then the subsystem
(\ref{ss2})--(\ref{ss4}) reduces to equations (\ref{short1}),
(\ref{short3}), while the subsystem (\ref{ss5})--(\ref{ss7}) becomes
equivalent to (\ref{short3}) (remember that $p_0=\p_0\theta$ and
$\textbf{p}=\nabla \theta)$.

Like (\ref{short1})--(\ref{short3}) the system with currents, i.e.,
(\ref{ss2})--(\ref{ss7}) is Poincar\'e invariant. The related
representation of the Lie algebra of Poincar\'e group can be
obtained by the prolongation of the basis elements
(\ref{kuriksha:yadro_operators}) to the first derivatives of
$\theta$ and adding analogous terms acting on components of the
current four-vector $j=(j_0,{\bf j})=(j_0, j_1,j_2, j_3)$:
\begin{eqnarray}\begin{aligned}&\hat P_0=\p_0, \quad \hat P_a=\p_{a},  \\
&\hat J_{ab}=x_a \p_{b}-x_b \p_{a}+B^a \p_{B^b}-B^b
\p_{B^a}+E^a\p_{E^b}-E^b
\p_{E^a}+p^a\p_{p^b}-p^b\p_{p^a}+j^a\p_{j^b}-j^b\p_{j^a},
\label{kuriksha:yadro_operatorsPROL}\\
&\hat J_{0a}=x_0 \p_{a}+x_a \p_0+\varepsilon_{abc} \left(E^b
\p_{B^c}-B^b \p_{E^c}\right)+p^0\p_{p^a} -p^a\p_{p^0}+j^0\p_{j^a}
-j^a\p_{j^0}.\end{aligned}
\end{eqnarray}
Generators (\ref{kuriksha:yadro_operatorsPROL}) do not include
differentials w.r.t. $j^4$ since this current component is not
changed under Lorentz transformations (i.e., it should be scalar).

 Being
applied to basis elements of algebra p(1,3) the IW contraction
consists of the transformation to a new basis
\begin{eqnarray}\hat J_{ab}\to J'_{ab}=\hat J_{ab}, \ \hat J_{0a}\to
J'_{0a}= \varepsilon \hat J_{0a},\ \hat P_0\to P'_0=\varepsilon^{-1}
\hat P_0,\ \hat P_a\to P'_a=\hat P_a\label{IW}\end{eqnarray} where
$\varepsilon$ is a small parameter equal to the inverse speed of
light. In fact we deal with the generalized IW contraction since
$P'_0$ is proportional to the inverse power of the small parameter.
In addition, the dependent and independent variables in
(\ref{kuriksha:yadro_operatorsPROL}) undergo the invertible
transformations $ E^a\to E'^a, B^a\to B'^a, p^\mu\to p'^\mu$,
$x^\mu\to x'^\mu$ where the primed quantities are functions of all
the unprimed ones and of $\varepsilon$. Moreover, the transformed
quantities should depend on the contracting parameter $\varepsilon$
in a tricky way, such that all transformed generators $P'_\mu,
J'_{ab}$ and $ J'_{0a}$ are kept nontrivial and nonsingular when
$\varepsilon\to 0$.

 The contractions of relativistic bi-vector fields
 (like  $\mathbf{E}, \mathbf{B}$) and four-vectors
 $p^\mu$ has been described in
  papers \cite{NN1} and \cite{NN2}.
  However, we need to contract simultaneously two subjects, i.e.,
  the basis elements of the Lorentz algebra given by equations
  (\ref{kuriksha:yadro_operatorsPROL}) and the system of equations
  (\ref{ss2})--(\ref{ss7}), which is a much
  more sophisticated problem.

  In order that transformation (\ref{IW}) be nonsingular in $\varepsilon$
  it should to be attended by the following transformations of the dependent
   and independent  variables:
  \begin{eqnarray}\begin{aligned}&
x'_0= t=\varepsilon{x_0}, \ \ x_a'=x_a,
\\
&{p_0'}={p_0},\ \ \textbf{p}'=\frac{\varepsilon}{2}
(\textbf{E}+\textbf{p}),\ \ \textbf{E}'={\varepsilon^{-1}}
(\textbf{E}-\textbf{p}),\ \ \textbf{B}'=\textbf{B},
\\&j'_0=\varepsilon^{-1}(j_0+j_4),\ \
j'_4=\frac\varepsilon2(j_0-j_4),\ \ \textbf{j}'=\textbf
{j}.\end{aligned}\label{5.5}
\end{eqnarray}

Making changes (\ref{5.5}) in equations
(\ref{kuriksha:yadro_operatorsPROL}), applying transformation
(\ref{IW}) and tending $\varepsilon\to 0$ we obtain the following
set of first order differential operators which for a basis of the
Galilei algebra:
\begin{eqnarray}\begin{aligned}& P'_0=\p_t, \quad  P'_a=\p_{a},  \\
& \begin{aligned}&J'_{ab}=x_a \p_{b}-x_b \p_{a}+B'^a \p_{B'^b}-B'^b
\p_{B'^a}+E'^a\p_{E'^b}-E'^b
\p_{E'^a}\\&+p'^a\p_{p'^b}-p'^b\p_{p'^a}+j'^a\p_{j'^b}-j'^b\p_{j'^a},\end{aligned}
\label{kuriksha:yadro_operatorsPROLA}\\
&J'_{0a}=t \p_{a}+\varepsilon_{abc} \left(p'^b \p_{B'^c}-B'^b
\p_{E'^c}\right)+p'^0\p_{E'^a}
-p'^a\p_{p'^0}-j'^a\p_{j'^0}.\end{aligned}
\end{eqnarray}

To obtain a consistent system of equations, we have to make changes
(\ref{5.5}) not directly in
 equations (\ref{ss2})--(\ref{ss7}),
 but in the equivalent system which includes equation (\ref{ss1}),
 (\ref{ss4}),
 (\ref{ss6})
 and  sums and half divergences of pairs of equation (\ref{ss2}),
 (\ref{ss5}) and  (\ref{ss3}), (\ref{ss7}).
 Then equating terms with lowest
  powers of $\varepsilon$ we obtain the following system:
   \beq\label{lastlast}\bea{l}\partial_t p'_0
+\nabla\cdot{\bf E}' +\kappa{\bf B}'
\cdot{\bf E}'=j'_0,\\
\partial_t {\bf p}'-\nabla\times{\bf B}'-\kappa(
p'_0{\bf B}'+{\bf p}'\times {\bf E}')={\bf j},\\\nabla\cdot{\bf
p}' -\kappa{\bf p}' \cdot{\bf B}'=j'_4 ,\\\nabla\cdot{\bf B }' =0,\\
\partial_t{\bf B}'+\nabla\times{\bf E}'=0,\\\
\partial_t{\bf p}'-\nabla p'_0=0, \ \
\nabla\times{\bf p}'=0
 \eea\eeq
and $p'^0=\p_t\theta', \  \mathbf{p}'=\nabla\theta'.$

Just equations (\ref{lastlast}) present the nonrelativistic limit of
system (\ref{ss2})--(\ref{ss7}). The Galilei invariance of system
(\ref{lastlast}) can be proven directly using the following
transformation laws which can be found by integrating the Lie
equations for generators (\ref{kuriksha:yadro_operatorsPROLA}):
\begin{eqnarray}\begin{aligned}&\mathbf{x}\to\mathbf{x}+\mathbf{v}t,
\ t\to t,\\& p'_0\to p'_0-{\bf v \cdot p'},\ \
 {\bf p}'\to{\bf p}',
 \ \ {\bf B}'={\bf B}'+{\bf v}\times{\bf p}',\\
 & {\bf E}'\to{\bf E}'-{\bf v}\times{\bf B}'+{\bf v}p'_0+
 {\bf v}({\bf v\cdot\bf p}')-\frac12{\bf v}^2{\bf p}'.\end{aligned}\label{SS}\end{eqnarray}

Thus we find the nonrelativistic limit for field equations of axion
electrodynamics with zero axion mass and nontrivial currents.
 It is interesting to note that  system (\ref{lastlast}) has
been discovered  in paper \cite{NN3}, starting with the requirement
of Galilei invariance for abstract vector fields. Thus our
contraction procedure presents a physical interpretation for the
Galilei invariant system for the indecomposable ten component field
deduced in \cite{NN3}, see equation (67) there. Namely, this Galilei
invariant system is nothing but a nonrelativistic limit of the field
equations of axion electrodynamics.

\section{Discussion}

The aim of the present paper is multifold. First we make group
classification of field equations of axion electrodynamics
(\ref{short1})--(\ref{short3}) which include an arbitrary function
$F$ depending on  $\theta$, and find the conservation laws generated
by these equations. Secondly, we find all continuous symmetries of
Chern-Simon electrodynamics and CFJ model which are closely related to axion electrodynamics. At the third place, we
discuss exact solutions for axion and e.m. fields
 which can be obtained using found symmetries. Finally,
  we define  a
correct nonrelativistic limit of the field equations of axion
electrodynamics.

As a result of the group classification we prove that the Poincar\'e
invariance is the maximal symmetry of the standard axion
electrodynamics and indicate the special forms of $F$ for which the
theory admits more extended symmetries. In accordance with
(\ref{kuriksha:rozsh_operators3}) such an extension appears for
trivial, constant and exponential interaction terms.

In spite of the transparent Poincar\'e invariance of the analyzed
equations, these results are nontrivial, since  possible  extensions
(\ref{kuriksha:rozsh_operators3}) of the basic  symmetries
 and the absence of other ones were
not evident a priori. Our results form group-theoretical grounds for
constructing of various axionic models. The detailed calculations of
Lie symmetries of equations (\ref{short1})--(\ref{short3}) are
presented in Appendix A.

In addition, in Appendix B we describe Lie symmetries of classical
electrodynamics modified by Chern-Simons term (i.e., symmetries of
the subsystem (\ref{short1}), (\ref{short3}) where $p_\mu$ is
treated as an external field), and symmetries of CFJ
electrodynamics. It is proven that the maximal continuous symmetry
of the Chern-Simons electrodynamics is given by the 17-parametrical
extended conformal group.

Conservation laws and the corresponding symmetries of the CFJ model
have been already studied in paper \cite{har}. However, the
symmetries which generate conservation laws are not the half of all
Lie symmetries \cite{olver}. Our research justifies and completes
the results of \cite{har}.  In particular we prove the completeness
of the list of space-time symmetries theory presented in paper
\cite{har}, and add this list by symmetry $D_1$ given by equation
(\ref{fin}).

Analyzing conservation laws caused by equations
(\ref{short1})--(\ref{short3}) we prove that  the energy-momentum
tensor does not depend on interaction between the electromagnetic
and axion fields and indicate that the additional symmetries
(\ref{kuriksha:rozsh_operators3}) do not generate conservation laws.
The first statement generalizes the observation present in book
\cite{obuh} to the case of arbitrary self interaction of  axionic
field.

Exact solutions of the field equations of axion electrodynamics are
discussed in Sections 5 and 6. There is a sufficiently large amount
of group solutions for these equations, but  we restrict ourselves
only to those which can be potentially important to physical
applications.

Analyzing the plane wave solutions we recognize the possibility of
the faster than light group velocities in axion electrodynamics.
However the corresponding energy velocities are subluminal and do
not lead to  causality violation.

Let us note that functions (\ref{wond1}) solve the field  equations
of CFJ electrodynamics which coincide with the system
(\ref{short1}), (\ref{short3}) where $p_\mu$ are constants. The
existence of faster than light solutions for these equations was
indicated in paper \cite{jackiw}, the enhanced discussion of the
plane wave solutions for (\ref{short1}), (\ref{short3}) can be found
in \cite{itin}. However, in  CFJ electrodynamics the energy is not
positive definite \cite{jackiw} while the energy density of axion
electrodynamics, discussed in section 5, is positive. In addition,
in the CFJ theory variables $p_\mu$ represent an external field
while in our approach they are dynamical variables satisfying
equations (\ref{short2}).

An important aspect of the presented exact solutions is that they
are requested in some problems of nonrelativistic and relativistic
quantum mechanics. In particular, as it was indicated in
\cite{ninni}, just the vectors of the electric and magnetic fields
described by relations (\ref{1/r3}) give rise to exactly solvable
Dirac equation for a charged particle anomalously interacting with
these fields. These fields are involved also into the
superintegrable nonrelativistic system discussed in \cite{N3}. The
electric field given by equation (\ref{Ea}) was used in \cite{N4} to
construct a new exactly solvable QM model with Fock symmetry. The
fields (\ref{1/r}) and (\ref{log}) are requested in superintegrable
models for particles with higher spins \cite{Pron}. Using the
formalism presented in \cite{NN4} it is possible to construct also
relativistic versions of these models.

Summarizing, the presented exact solutions are not nice mathematical
toys only. In contrary, they have a nontrivial physical content.
These solutions  give rise to exactly solvable QM systems enumerated
in the previous paragraph, which in principle can be used to predict
physically verifiable effects. That opens  ways  for  finding new
arguments for the real existence of the pseudoscalar axion field.

Solutions discussed in Section 4 represent only a part of exact
solutions which can be obtained using three dimensional subalgebras
of algebra p(1,3). The complete list of them can be found in
\cite{kura}. Notice that among these solutions there are rather general
ones which include six arbitrary functions \cite{kura}.

 A special goal of this paper was to
present a correct nonrelativistic limit of equations of axion
electrodynamics. To achieve this goal we use the generalized IW
contraction of the corresponding representation of the Poincar\'e
group. As a result we prove that the limiting case of these
equations is nothing but the Galilei-invariant system for the
ten-component vector field obtained earlier in paper \cite{NN3}.
This result casts light on the physical content of the model
discussed in \cite{NN3}. In addition, the  contraction procedure can
be used for transforming solutions discussed in Sections 4, 5 (and
other solutions found in \cite{kura}) to solutions of system
(\ref{lastlast}).

Some results of this paper had been announced in conferences, see
Proceedings \cite{NNK}.

\vspace{2mm}

\begin{center} ACKNOWLEDGEMENTS \end{center}

\vspace{1mm}

This work was partially supported by research program
"Cosmomicrophysics" of the National Academy of Sciences of Ukraine
(state registration  number 0109U003207).

\appendix
\section{LIE SYMMETRIES of equations
(\ref{short1})--(\ref{short3})}

The group analysis of differential equations is a nice field of
mathematics whose fundamentals had been created by Sophus Lie as
long ago as in the end of the eighteenth century. The foundations of
the group analysis are expounded in many books, the most popular of
which is monograph \cite{olver}. Nevertheless we will present a very
short sketch of the Lie algorithm addressed  to more physically
oriented readers.

The basic idea of the classical Lie algorithm is to treat a
differential equation (or a system of equations) as a manifold in
the multidimensional Euclidean space whose basis elements include
the dependent and independent variables and also the  derivatives
 present in the equation. In our case we have four independent variables $x^0,
 x^1, x^2, x^3$, seven dependent variables $F^{\mu\nu}, \theta$ and
 38 differential variables $F^{\mu\nu}_{,\lambda}=
 \p_\lambda F^{\mu\nu},\ \theta_{,\nu}=\p_\nu \theta$ and
 $\theta_{,\mu\nu}=\p_\mu\p_\nu\theta$. Equations
 (\ref{short1})--(\ref{short3}) define a hypersurface $\cal F$ in the
 Euclidean space $\Re_{49}$ with variables
 $x^\mu,F^{\mu\nu},\ \theta,\ F^{\mu\nu}_{,\lambda},\ \theta_{,\nu},
 \theta_{,\mu\nu}$. Moreover, $\cal F$ is a smooth
 manifold in $\Re_{49}$.

  Local continuous
symmetries of system (\ref{short1})--(\ref{short3}) can be treated
as continuous transformations in $\Re_{49}$ which keep this manifold
invariant.  To find these symmetries it is possible to imply the
infinitesimal invariance criterium, i.e., consider transformations
close to the identity one:
\begin{eqnarray}x^\mu\to x'^\mu=x^\mu+\varepsilon \xi^\mu,\quad F^{\mu\nu}\to
F'^{\mu\nu}=F^{\mu\nu}+\varepsilon \eta^{\mu\nu}, \quad
\theta\to\theta'=\theta+\varepsilon\sigma\label{Dur1}\end{eqnarray}
where $\varepsilon$ is a
 small transformation parameter and
$\xi^\mu,\ \eta^{\mu\nu},\ \sigma$ are some functions of $x^\mu,
F^{\mu\nu}$ and $\theta$, and ask for the form invariance of
equations (\ref{short1})--(\ref{short3}) w.r.t. the change of
variables (\ref{Dur1}). Transformations (\ref{Dur1}) can be formally
represented as
\begin{gather*}f\to f'=(1+\varepsilon Q)f\end{gather*}
where $f$ is a vector whose components are independent variables
$x^\mu$ and dependent variables $F^{\mu\nu}, \ \theta$, and $Q$ is
the infinitesimal operator:
\begin{eqnarray}
Q= \xi^\mu\p_{\mu}+\frac12 \eta^{\mu\nu} \p_{F^{\mu\nu}}+\sigma
\p_{\theta}.\label{kuriksha:operator}
\end{eqnarray}

 Starting with
(\ref{Dur1}) we can find transformations for the differential
variables:
\begin{eqnarray}\label{dur2}F^{\mu\nu}_{,\lambda}\to
F'^{\mu\nu}_{,\lambda}= F^{\mu\nu}_{,\lambda}+\varepsilon
\eta^{\mu\nu}_{\lambda},\quad
\theta_{,\mu}\to\theta'_{,\mu}=\theta_{,\mu}+\varepsilon
\sigma_\mu,\quad
\theta_{,\mu\nu}\to\theta'_{,\mu\nu}=\theta_{,\mu\nu}+\varepsilon
\sigma_{\mu\nu}\end{eqnarray} where
\begin{gather*}
\eta_{\mu}^{\nu\sigma}=D_{\mu}(\eta^{\nu\sigma})-F^{\nu\sigma}_{,\lambda}
D_{\mu}(\xi^\lambda),\ \ \
\sigma_{\mu}=D_{\mu}(\sigma)-\theta_{,\lambda}D_{\mu}(\xi^\lambda),
\\ \sigma_{\nu\mu}=D_{\nu}(\sigma_{\mu})-\theta_{,\mu
\lambda}D_{\nu}(\xi^\lambda)
\end{gather*}
and $D_{\mu}=\p_\mu+\frac12F^{\nu\sigma}_{,\mu}\p_{F^{\nu\sigma}}+
\theta_{,\mu}\p_{\theta}+ \theta_{,\nu \mu}\p_{\theta_{,\nu}}.$

Using (\ref{dur2}), the invariance condition  for system
(\ref{short1})--(\ref{short3}) can be written in the following form:
\begin{eqnarray}Q_{(2)}{\cal F}|_{{\cal F}=0}=0
\end{eqnarray}
where $Q_{(2)}$ is  the infinitesimal operator
(\ref{kuriksha:operator}) prolonged to the first and second
derivatives:
\begin{eqnarray}\label{5}Q_{(2)}= Q+\frac12\eta_{\mu}^{\nu\sigma} \p_{
F^{\nu\sigma}_{,\mu}}+ \sigma_{\mu}
\p_{\theta_{,\mu}}+\frac12\sigma_{\mu\nu} \p_{\theta_{,\mu \nu}}
\end{eqnarray} and  $\cal F$
is the manifold defined by relations (\ref{short1})--(\ref{short3}).

Acting by operator (\ref{5}) on differential forms
(\ref{short1})--(\ref{short3}) and equating coefficients for
linearly independent functions $F^{\mu\nu},\theta$ and their
derivatives we obtain the following system of determining equations
for the coefficients $\xi^{\mu},\ \eta^{\mu\nu}$ and $\sigma$:
 \begin{eqnarray}&&
\xi^\mu_{F^{\nu\lambda}}=0,  \quad \xi^\mu_{\theta}=0, \quad
\xi^\mu_{,\nu}+\xi^\nu_{,\mu}=\frac12\delta^\nu_\mu\xi^\sigma_{,\sigma},
\label{kuriksha:det_eq}\\&& \sigma_{F^{\mu\nu}}=0, \quad
\sigma_{\theta\theta}=0, \label{kuriksha:det_eq_2}
\\&&
\p_\mu \p^\mu
\sigma+\left(\sigma_{\theta}-\frac12\xi^\mu_{,\mu}\right)
(F+\frac{\kappa}2F_{\mu\nu}\widetilde
F^{\mu\nu})-\frac\kappa2\eta^{\mu\nu}\widetilde F_{\mu\nu}-\sigma
F_\theta=0, \label{kuriksha:det_eq_4}\\&&
\begin{aligned}&
\eta^{\mu\nu}_{,\nu}=\widetilde F^{\mu\nu}\sigma_{,\nu},\quad
{\varepsilon^{\mu\nu}}_
{\alpha\sigma}\eta^{\alpha\sigma}_{,\nu}=0,\quad 2\sigma_{\theta
,\mu}=\p_\nu\p^\nu \xi_\mu,\\&\eta^{\mu\nu}-
\frac12{\varepsilon^{\mu\nu}}_{\alpha\sigma}\eta^{\alpha\sigma}_\theta
+F^{\mu\nu}\sigma_\theta
+F^{\nu\alpha}\xi^\mu_{,\alpha}-F^{\mu\alpha}\xi^\nu_{,\alpha}=0,
\\
& \xi^\mu_{,\nu}+\eta^{\lambda\nu}_{F^{\lambda\mu}}=0,
\quad\eta^{\lambda\kappa}_{F^{\lambda\kappa}}=
\eta^{\mu\nu}_{F^{\mu\nu}},\quad \eta^{\mu\nu}_\theta= F^{\mu\nu}
\eta^{\lambda\kappa}_ {\widetilde
F^{\lambda\kappa}}\end{aligned}\label{kuriksha:det_eq_1}
\end{eqnarray}
where the subscripts $F_{\nu\lambda}$ and $\theta$ denote the
derivatives with respect to the corresponding variables:
$\xi^\mu_{F^{\nu\lambda}}=\frac{\p \xi^\mu}{\p F^{\nu\lambda}}, \
\eta^{\mu\nu}_\theta=\frac{\partial \eta^{\mu\nu}}{\p \theta}$,
etc., $\delta^\nu_\mu$ is the Kronecker symbol and there are no sums
over the repeating indices in the last line of equation
(\ref{kuriksha:det_eq_1}).

In accordance with equations (\ref{kuriksha:det_eq}) functions
$\xi^\mu$ do not depend on $F^{\mu\nu}$ and $ \theta$ and, moreover,
they are Killing vectors in the space of independent variables:
\beq\label{KV}\xi^\mu=2x^\mu f^\nu  x_\nu-f^\mu x_\nu
x^\nu+c^{\mu\nu}x_\nu+d x^\mu+e^\mu \eeq where $f^\mu$, $d$, $e^\mu$
and  $c^{\mu\nu}=-c^{\nu\mu}$ are arbitrary constants.

It follows from (\ref{kuriksha:det_eq_2}) that $
\sigma=\varphi_1\theta+\varphi_2, $ where $\varphi_1$ and
$\varphi_2$ are functions of $x_\mu$. Substituting this expression
into (\ref{kuriksha:det_eq_4}) we obtain the following equation:
\begin{eqnarray}\begin{aligned}
&\varphi_1 \theta {F_\theta}+\varphi_2{F_\theta}+\left(2\xi^0_{,0}-
\varphi_1\right)\left(F+\frac\kappa2F_{\mu\nu}
\widetilde F^{\mu\nu}\right)\\
&+ \kappa \eta^{\mu\nu}\widetilde F_{\mu\nu}-\theta\p_\alpha
\p^\alpha \varphi_1-\p_\alpha \p^\alpha
\varphi_2-2p^\mu\p_\mu\varphi_1=0.\end{aligned}\label{kuriksha:det_eq_5}
\end{eqnarray}

Let the terms $\theta {F}_\theta,\ {F}_\theta, \ F,\text{ and } 1$
be linearly independent. Then it follows from
(\ref{kuriksha:det_eq_5}) and (\ref{KV}) that
$\varphi_1=\varphi_2=0, f^\nu=0$ and $\eta^{\mu\nu}\widetilde
F_{\mu\nu}=0$. Thus, using (\ref{kuriksha:det_eq_1}) and (\ref{KV})
we obtain:
 \begin{eqnarray}\label{000}
\xi^\mu=c^{\mu\nu}x_\nu+e^\mu,
 \
\ \ \eta^{\mu\nu}=c^{\mu\alpha}{F^\nu}_\alpha-c^{\nu\alpha}
{F^\mu}_\alpha,\ \ \sigma=0.
\end{eqnarray}

Substituting (\ref{000}) into (\ref{kuriksha:operator}) we receive a
linear combination of infinitesimal operators
(\ref{kuriksha:yadro_operators}) which form a basis of the Lie
algebra of Poincar\'e group P(1,3).
 Thus {\it the group
P(1,3) is the maximal continuous symmetry group of system
(\ref{short1})--(\ref{short3}) with {\it arbitrary} function}
$F(\theta)$.

The possible extensions of this symmetry which appears for some
particular functions $F$ are enumerated in equations (11).
\section{Symmetries of Chern-Simons and Carrol-Field-Jackiw models}
\renewcommand{\theequation}{B\arabic{equation}} \setcounter{equation}{0}

Let us discuss symmetries of two models which are closely related to
the axion electrodynamics. The first is the classical
electrodynamics modified by adding the  Chern-Simons terms. It is
based on field equation (\ref{short1}), (\ref{short3}), but does not
include the dynamical equation for $p_\mu$ which is treated as an
external field.

Let us suppose that vector $p_\mu$ can be presented as a
four-gradient of some scalar function $\theta$. Then, repeating the
procedure used above we again come to the determining equations
(\ref{kuriksha:det_eq}), (\ref{kuriksha:det_eq_2}),
(\ref{kuriksha:det_eq_1}) while equation (\ref{kuriksha:det_eq_2})
would be absent. Solving these determining equations we obtain the
17-dimensional symmetry algebra whose basis elements are given by
equations (\ref{kuriksha:yadro_operators}) and by the following
equations:
\begin{eqnarray}&&\begin{aligned}&D_1=F^{\mu\nu}\p_{F^{\mu\nu}},\quad
D_2=x^\mu\p_\mu-D_1,\\& K_\mu=2x_\mu D_2-x_\nu x^\nu \p_\mu +2x^\nu
S_{\mu\nu},\end{aligned}\label{Kmu}\\&&\quad P_4=\frac\p{\p
\theta}\label{DP}\end{eqnarray}where $S_{\mu\nu}$ are operators
defined by equation (\ref{smunu}).

Operators (\ref{kuriksha:yadro_operators}) and (\ref{Kmu}) form a
basis of the Lie algebra of  conformal group, while operators $P_4$
and $D_2$ belong to the central extension of this algebra.

Thus the field equations of Chern-Simons electrodynamics are
invariant w.r.t. the conformal group and admit two additional
symmetries, i.e., scaling of the strength tensor of electromagnetic
field  and shifts of function $\theta$.

If vector field $p_\mu$ is not supposed a priori to be a
four-gradient, we come to the 16-dimensional symmetry algebra of
equations (\ref{short1}), (\ref{short3}). The explicit form of its
basis elements can be obtained from (\ref{kuriksha:yadro_operators})
and (\ref{Kmu}) by the change $S_{\mu\nu}\to \tilde S_{\mu\nu}$ and
$D_2\to\tilde D_2$ where
\begin{gather*}\tilde
S_{\mu\nu}=S_{\mu\nu}+p_\mu\p_{p^\nu}-p_\nu\p_{p^\mu},\quad \tilde
D_2=D_2-p_\mu\p_{p^\mu}.\end{gather*}

The second model which we consider here is the  Carrol-Field-Jackiw
one \cite{jackiw}. This is a particular version of the Chern-Simons
electrodynamics corresponding to a constant vector $p_\mu$.
Considering equations (\ref{short1}), (\ref{short3}) with the
additional condition \beq \p_\nu p_\mu=0,\label{acond}\eeq we
conclude that in this case the  symmetry algebra is reduced to
$\tilde{\text{p}}(1,3)\oplus \text{A}_1$ where
$\tilde{\text{p}}(1,3)$ is the extended Poincar\'e algebra whose
basis elements are operators (\ref{kuriksha:yadro_operators}) and
operator $\tilde D_2$, while  $\text{A}_1$ is the one-dimensional
algebra spanned on $D_1$. The reason of this reduction is that
equations (\ref{acond}) are not invariant w.r.t the conformal
transformations generated by operators $K_\mu$.

Thus the system  (\ref{short1}), (\ref{short3}), (\ref{acond}) is
not invariant w.r.t. the conformal transformations, but it is still
invariant w.r.t. the extended Poincar\'e group $\tilde{
\text{P}}(1,3)$ which includes the ordinary Poincar\'e group and
accordant scalings of dependent and independent variables. Equations
(\ref{acond}) require that $p_\mu$ are (arbitrary) constants. If
these constants are fixed, the symmetry group $\tilde{
\text{P}}(1,3)$ is reduced to its subgroup which keeps them
invariant. Namely, for time-like, space-like and light-like vectors
$p=(p_0, p_1, p_2, p_3)$ we can set
\begin{eqnarray}p=(\mu,0,0,0),\label{p1}\\ p=(0, 0, 0,
\mu),\label{p2}\end{eqnarray} and
\begin{eqnarray}p=(\mu,0,0,\mu)\label{p3}\end{eqnarray} correspondingly. Then
the subgroups of $\tilde{ \text{P}}(1,3)$ which keep (\ref{p1}),
(\ref{p2}) or (\ref{p3}) invariant are the Euclid group E(3), the
Poincar\'e group P(1,2) in (1+2)-dimensional space or the extended
Galilei group $\tilde {\text{G}}(1,2)$ respectively. The basis
elements of the corresponding Lie algebras are
\begin{eqnarray}\begin{aligned}&\langle P_1, P_2, P_3, J_{12}, J_{23},
J_{31}\rangle,\\& \langle P_1, P_2, P_3, J_{12}, J_{01},
J_{02}\rangle\end{aligned}\label{lla}\end{eqnarray} or
\begin{eqnarray}\label{llb} \langle P_1, P_2, P_3, J_{12},
G_1=J_{01}+J_{31}, G_2=J_{02}+J_{32}, D_3=\tilde
D_2+J_{03}\rangle\end{eqnarray} where $P_a, J_{\mu\nu}$ and $ D_2$
are generators given in (\ref{kuriksha:yadro_operators}) and
(\ref{Kmu}). In addition, in all cases (\ref{p1})--(\ref{p3}) there
are additional symmetries \begin{eqnarray}P_0=\frac{\p}{\p x_0}\label{defin}\end{eqnarray}
and
\begin{eqnarray}D_1=F^{\mu\nu}\p_{F^{\mu\nu}}.\label{fin}\end{eqnarray}

The list of symmetries (\ref{lla}) -- (\ref{llb}) had been found in
paper \cite{har} starting with conservation laws for equations
(\ref{short1}) and (\ref{short3}). Our approach is more
straightforward and guarantees finding of all Lie symmetries
including nonvariational ones. An example of such nonvariational
symmetry which cannot be found with approach used in \cite{har} is
the dilatation whose generator $D_1$ is given by equation
(\ref{fin}).


\begin{thebibliography}{99}

\bibitem{pec}R. D. Peccei and H. R. Quinn,  Phys. Rev. Lett. {\bf 38},
1440 (1977).

\bibitem{weinberg}S. Weinberg,  Phys. Rev. Lett. {\bf 40}, 223
(1978).

\bibitem{wilczek1} F. Wilczek,  Phys. Rev. Lett. {\bf 40}, 279 (1978).

\bibitem{wilczek} F. Wilczek,  Phys. Rev. Lett. {\bf 58}, 1799 (1987).

\bibitem{Ni} W.-T. Ni, Bull.
Am. Phys. Soc. {\bf 19}, 655 (1974).

\bibitem{raflet} G. G. Raffelt,  Phys. Rep. {\bf 198}, 1 (1990).

\bibitem{Qi} X-L. Qi, T. L. Hughes, and S-C. Zhang,  Phys. Rev. B {\bf  78},
 195424 (2008).

\bibitem{heht}F. W. Hehl, Y.N. Obukhov J.-P. Rivera and H. Schmid,
 Eur. Phys. J. B {\bf  71}, 321–329 (2009)

\bibitem{jackiw} S. M. Carroll, G. B. Field and R. Jackiw,
 Phys. Rev D {\bf 41} 1231 (1990).

\bibitem{chern} S.S. Chern  and  J. Simons,  Annals Math. {\bf 99} 48 (1974).

 \bibitem{ninni} E. Ferraro, A. Messina and A.G. Nikitin,
 Phys. Rev. A {\bf 81}, 042108 (2010).

 \bibitem{N4} A. G. Nikitin,  arXiv:1205.3094  (2012).


 \bibitem{Nik2}A. G. Nikitin and Y. Karadzhov,  J. Phys. A: 44 (2011) 305204.

 \bibitem{NKa2}A. G. Nikitin and Y. Karadzhov,  J. Phys. A: 44 (2011)
 445202.

 \bibitem{N3} A.G. Nikitin, arXiv:1204.5902v2 (2012).

 \bibitem{NN3}J. Niederle and A.G. Nikitin,  J. Phys. A: Math. Theor. { 42} 105207
(2009).

\bibitem{har}A.J. Hariton and R. Lehnert,  Phys. Lett. A {\bf 367}
 11 (2007).

 \bibitem{Pron} A. G. Nikitin, J. Phys. A: Math. Theor. 45 (2012) 225205.

   \bibitem{lebellac} M. Le Bellac  and J.-M. L\'evy-Leblond, Nuovo Cimento
B 14, 217 (1973).

\bibitem{Hol} P. Holland  and H.R. Brown,
 Studies in History and Philosophy of Science  { 34}, 161,
(2003).

\bibitem{contraction} E. In\"on\"u  and E.P. Wigner, Proc. Nat. Acad. Sci. US
{ 39} 510 (1953).

\bibitem{olver} P. Olver,  {\it Application of Lie groups to
Differential
 equations}  (Second Edition, Springer-Verlag, New York, 2000),  electronic version:  PJ Olver - 2000 - books.google.com

 \bibitem{kura} Oksana Kuriksha and A.G. Nikitin, arXiv:1002.0064v4 (2012),
 http://dx.doi.org/10.1016/j.cnsns.2012.04.009 (2012).

 \bibitem{patera}J. Patera, P. Winternitz,
and H. Zassenhaus, J. Math. Phys. { 16}, 1597 (1975).

  \bibitem{Bril} L. Brillouin, Wave Propagation and Group Velocity (Academic,
New York, 1960).

 \bibitem{smith} R. L. Smith,  Am. J. Phys. 38, 978-983 (1970).

\bibitem{bloch} S. C. Bloch,  Am. J. Phys.
45, 538 (1977).

\bibitem{NN1} M. de Montigny,  J. Niederle  and A.G. Nikitin,
 J. Phys. A: Math. Theor. { 39} 9365 (2006).

\bibitem{NN2}  J. Niederle  and A.G. Nikitin,    Czech.
     J. Phys. { 56} 1243 (2006).

\bibitem{obuh}F.W. Hehl and Y.N. Obukhov,
 {\it Foundations of Classical Electrodynamics -- Charge, Flux and Metric}
 (Birkhauser 2003).


\bibitem{itin} Y. Itin, Phys. Rev. D { 76} 087505 (2007);\  \ Y. Itin.
Gen. Rel. Grav. { 40} 1219 (2008).

 \bibitem{NN4} J. Niederle and A. G. Nikitin, Phys. Rev. D
 { 64} 125013 (2001).

 \bibitem{NNK} J. Niederle, A. G. Nikitin and O. Kuriksha, In: Proceedings of the Fifth International Workshop
     "Group Analysys of Differential Equations and Integrable Systems",
      June 6-10, 2010, Protaras, Cyprus, University of Cyprus, Nikosia, 2011,
      pp. 152-163;\\J. Niederle, A. G. Nikitin and O. Kuriksha, Acta Polytechnica {\bf 50} 96 (2010).
\end{thebibliography}
\end{document}